\documentclass[twocolumn,aps,floats,showpacs,amsmath]{revtex4}
\usepackage{amssymb,amsmath}
\usepackage{graphicx,float}

\begin{document}
\title{ \Large \bf  Modelling conflicts with cluster dynamics on networks}
\author{\large  \flushleft 
Bosiljka Tadi\'c$^{1}$  and G.J.Rodgers$^2$
} 

\affiliation{ 
$^1$Department  for Theoretical Physics; Jo\v{z}ef Stefan Institute; 
P.O. Box 3000; SI-1001 Ljubljana; Slovenia, 
$^2$Department of Mathematical Sciences;
Brunel University;
Uxbridge
Middlesex
UB8 3PH
United Kingdom
\\ \hspace{1cm}} 


\begin{abstract}
We introduce  cluster dynamical models of conflicts in which only 
the largest cluster can be involved in an action. This mimics the 
situations in which an attack is planned by a central body, and the largest attack 
force is used. We study the model in its annealed random graph version, on a fixed 
network, and on a network evolving through the actions. The sizes of actions are distributed with a 
power-law tail, however, the exponent is non-universal and depends on the frequency 
of actions and sparseness of the available connections between units. Allowing the 
network reconstruction over time in a self-organized manner, e.g., by adding the 
links based on previous liaisons between units, we find that the power-law exponent 
depends on the evolution time of the network. Its lower limit is given by 
the universal value 5/2, derived analytically for the case of random fragmentation processes. 
In the temporal patterns behind the size of actions we find long-range correlations in the time series of 
number of clusters and non-trivial distribution of time that a unit waits between two actions. 
 In the case of  an evolving network the distribution develops 
a power-law tail, indicating that through the repeated actions, the 
system develops internal structure which is not just more 
effective in terms of the size of events, but also has a  full hierarchy of units.
\end{abstract}
\maketitle
\section{Introduction}

The study of coagulation-fragmentation models in physics have a long history,
and have been used to explain a variety of physical phenomena, including the 
formation of aerosols, colloidal aggregates, polymers and celestial bodies \cite{wattis}. 
More recently the basic theory, originally developed by Smoluchowski \cite{smol} 
and Becker-Doring \cite{becker} 
has been adapted to model group behaviour in human-dynamics. 
		The first microscopic model of herding in financial markets consisted
 of randomly connected agents \cite{bouchaud}. In this model agents are connected 
with probability $a$ and disconnected with probability $1-a$. Agents that are in the same 
group share information and make the same decisions in the market. The 
parameter $a$ was tuned to the percolation threshold to obtain a power-law 
distribution of group sizes and hence a power-law distribution of returns. 
	
	Inspired by this, in 2000 Eguiluz and Zimmermann \cite{EZ} (EZ) introduced a 
coagulation-fragmentation model of herding in financial markets. The EZ 
model is a kinetic version of the model in \cite{bouchaud}, in which the group sizes 
emerge more naturally in the limit of large time. At each time step with 
probability $p$ an edge is introduced between two randomly selected 
agents or with probability $1-p$ the group of a randomly selected agent is fragmented.
	In \cite{geoff1} Dhulst and Rodgers solved this model exactly and showed 
that in the limit $t \rightarrow \infty$ the system evolves to a stationary cluster 
size distribution  with the number of clusters of size $s$, $n_s$, is 
given by
\begin{equation}
n_s = \frac{p^{s-1}(2s-2)!}{(p+1)^{2s-1}s!^2}N
\label{eq-ns}
\end{equation}
	where N is the number of agents in the system. In the limit 
$p \rightarrow 1$ this result can be expanded for large $s$ to give a power-law 
with $n_s \sim s^{-\tau}$ and $\tau = 5/2$. 
	
	This basic model has been used \cite{johnson} to explain the apparent 
ubiquity of the exponent $\tau = 5/2$ in the data from modern insurgent 
warfare. Many authors have studied wars and conflicts empirically, and many have 
concluded that the distribution of daily casualties, or of casualties per attack, 
has a 
power-law distribution \cite{richardson48,richardson60}. A wide range of 
exponents have been reported.
 In \cite{newman} an exponent of  $\tau =1.8$ is reported for the intensity of 
119 old wars between 1816 and 1980. 
Casualty numbers in global terrorist attacks, since 1968, have 
$\tau= 1.7$ for G7 countries and $\tau = 2.5$ for non-G7 countries \cite{clauset}. More recently, 
in \cite{johnson} the daily data from killings and injuries in Columbia and civilian 
casualties in Iraq were examined. The total data sets gave good power-law
 distributions with $\tau = 3$ for Columbia and $\tau = 2$ for Iraq. By 
reducing the time window within these data sets, and then sliding the time
 window forward, the authors of \cite{johnson} were able to calculate a series for 
$\tau$ as a function of time. This revealed that as time window is shifted towards recent time the 
exponent $\tau$ in both the Columbian and Iraqi datasets was tending to 
the value $\tau = 5/2$. This caused the authors of \cite{johnson} to develop a new 
model of modern insurgent warfare based on a dynamical model of herding 
introduced in \cite{EZ,geoff1}.

In financial markets, the return is 
equal to the difference in the number of buyers and sellers at a particular time, 
or alternatively to the supply and demand balance at a particular time \cite{rose}. 
Hence in 
the models of financial markets the probability of having a return of size $s$ is 
proportional to $sn_s$. In models of wars and conflicts the units can be thought 
of as troops, weapons or equipment and it is an assumption in these models 
\cite{johnson} that the number of casualties inflicted by a force of strength $s$ is 
proportional to $s$.
In these models the parameter $p$ is an {\it external} fixed 
parameter, however, in the real dynamics this probability might emerge from
another stochastic process, inherently related to the main 
dynamics. One would expect that the stochastic processes would be quite 
different in the market and conflict dynamical systems.

The empirical studies of war and terrorist actions demonstrate 
\cite{johnson,richardson48,richardson60,newman} that the power-law behaviour 
is {\it non-universal}, 
with the exponent $\tau$ depending on the type of conflict (insurgence, 
guerrilla, terrorist, etc...)
 and on the geographical location and period of time considered.
In contrast the original model with 
the coagulation-fragmentation of random clusters always leads to a universal 
exponent $\tau=5/2$. Thus the EZ model is not able to capture the underlying 
mechanisms 
which lead to the variety of observed power-laws. Moreover, the full fragmentation 
of clusters, which is suitable for herding in market dynamics, might not be 
appropriate for the conflicts.

In this paper we study new models of conflicts with the  cluster  dynamics on 
networks when only the largest cluster can be involved in an action. This mimics the 
situations in which an attack is planned by a central body, and the largest attack 
force is used, and where the units interact within a social or technological
network.   Following the action,  the  network is either reset back to the original
 structure (fixed network model), or evolves by adding new links between units 
involved in the action (emergent network model).  The evolving network model provides a mechanism 
for understanding the power-law exponent dependence on the evolution time, similar to the empirical data.
Furthermore, in our simulations we analyse the temporal behavior of each unit, 
which is behind the emergent power-laws. This allows us to unravel the changing 
nature of the group actions and monitor emergent structure of their connections. For 
comparison, we also study  the cluster dynamics with maximal cluster fragmentation 
on an annealed random graph structure, which is closer to the original model and 
the situation in real market dynamics.

In Section \ref{sec-model} we introduce the microscopic rules of the model and 
present the simulations on the annealed random graph, the situation which 
parallels the original coagulation-fragmentation model of market crashes. 
We also define and compute the temporal quantities that can be monitored in real 
data. In section \ref{sec-networks} we present simulation results corresponding 
to the cluster dynamics on a network with  fixed and evolving network topology. 
Finally, in section \ref{sec-conclusions} we give a brief summary of the results 
and a discussion, in particular on the mechanisms revealed within our model, which 
might drive the evolution of modern conflicts.

\section{Cluster-aggregation and largest cluster fragmentation with an annealed
 random graph topology\label{sec-model}}
In this section we consider a minimal extension of the original 
aggregation--fragmentation model \cite{EZ,geoff1},  
in which instead of choosing a random group to fragment 
we always choose the largest group. 
At each time step with probability $p$ a pair of units is selected and 
joined together, whereas with probability $1-p$ the 
largest cluster is found and fragmented. As in the original model of market dynamics
 \cite{EZ,geoff1}, in the system of $N$ units any randomly selected pair can join 
each 
other in the aggregation event, whereas all the links between pairs within the 
cluster are removed when the cluster is destroyed. In terms of topology (see later) 
this situation is represented by a random graph with an annealed link structure. 
The effects of a fixed and evolving network underlying the aggregation 
processes will be studied in  section \ref{sec-networks}.

\subsection{Cluster size statistics and temporal structure}
\begin{widetext}
\begin{figure*}
\begin{tabular}{cc} 
{\large (a)}&{\large (b)}\\
\resizebox{20.4pc}{!}{\includegraphics{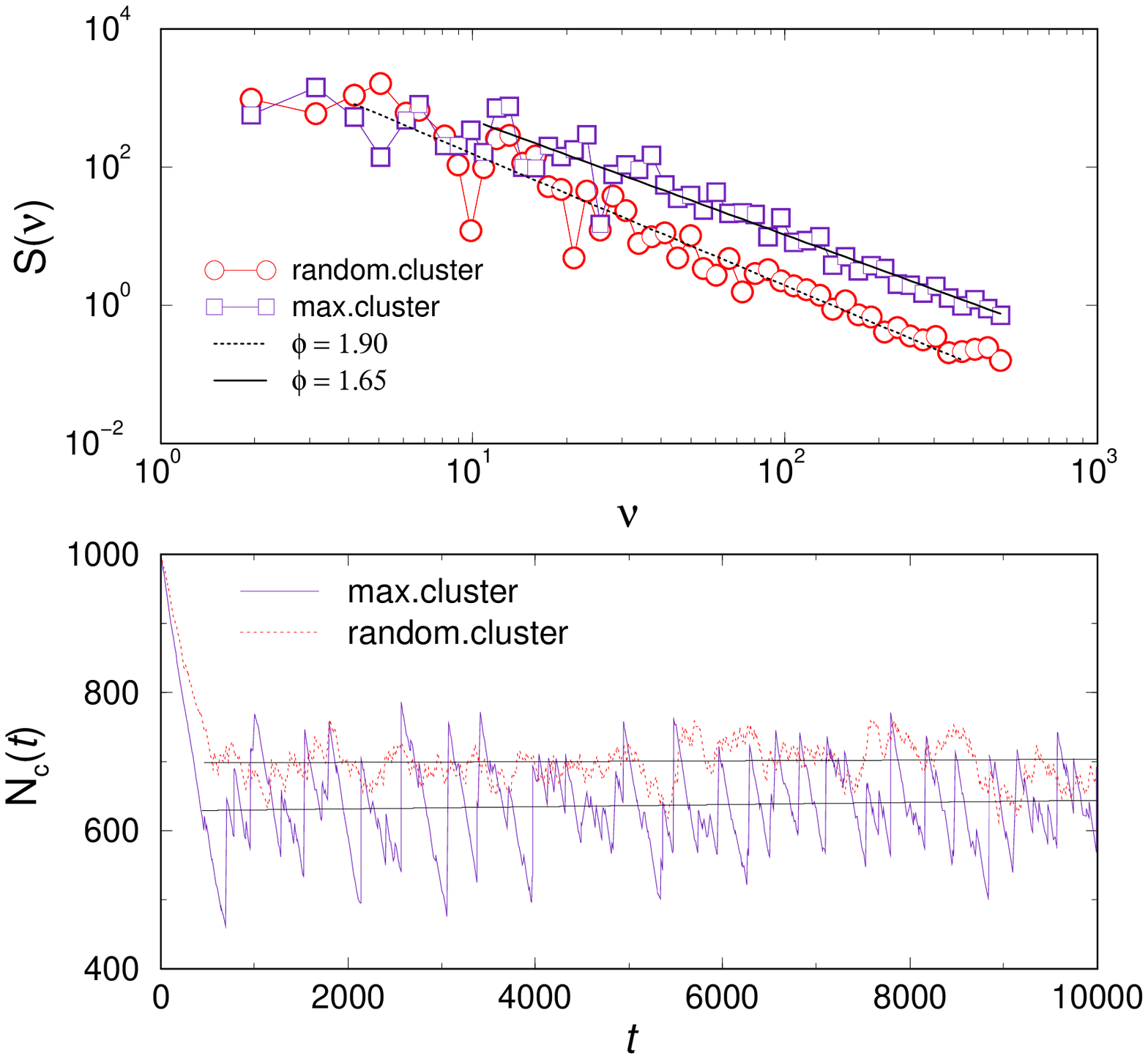}} &
\resizebox{20.4pc}{!}{\includegraphics{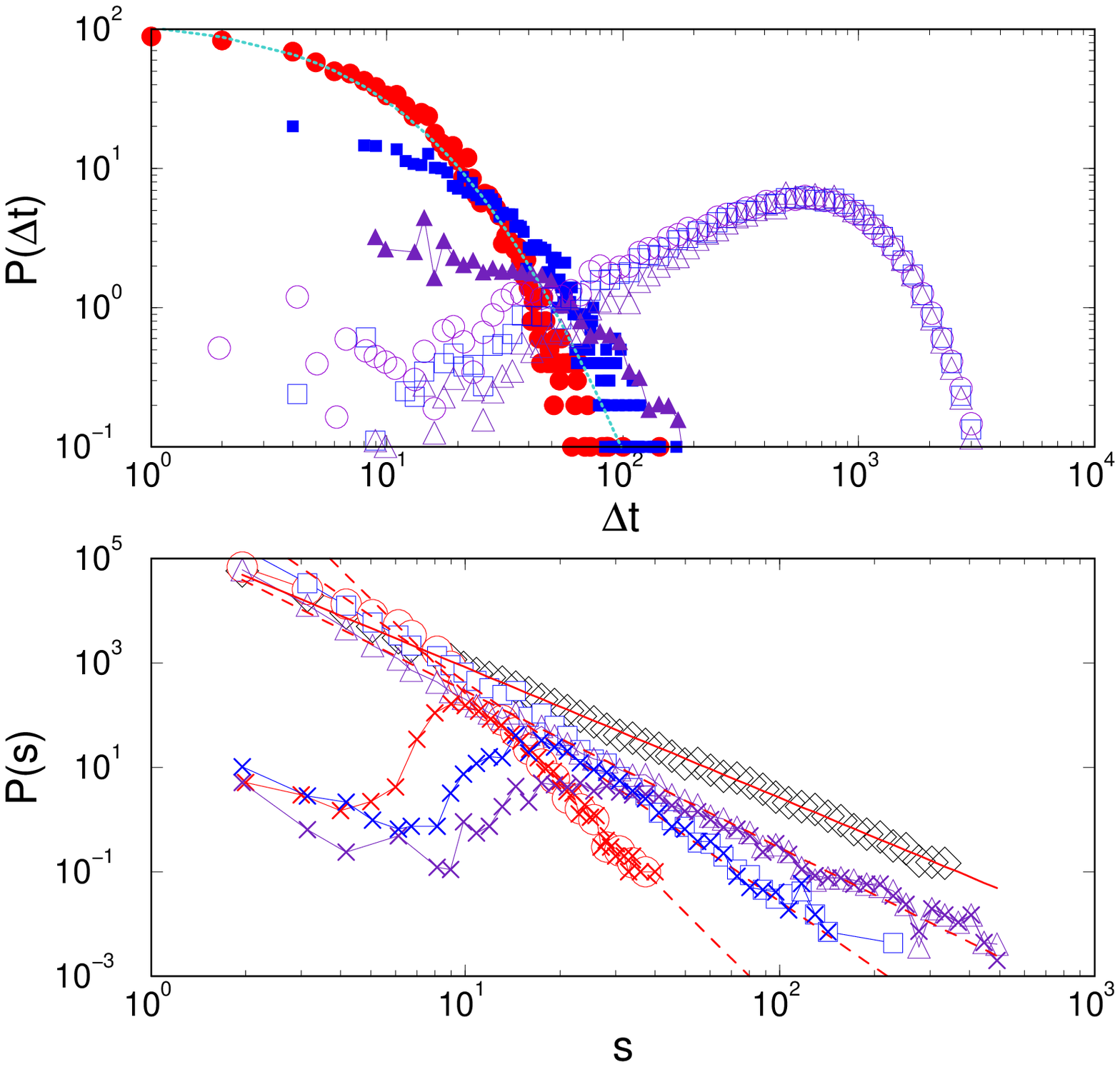}}\\ 
{\large (c)}&{\large (d)}\\
\end{tabular}
\caption{For the annealed random-graph topology: Time series (c) and their power spectra (a) of fluctuating 
number of clusters for aggregation probability $p=0.9$ and two crush mechanisms---maximal cluster crush and random cluster crush.
(d) Distribution $P(s)$ of size $s$ of all clusters, empty symbols, and size  of crushed maximal clusters, crosses, for three values of parameter $p=$ 0.9,0.95, and 0.98.
  Dashed lines indicate slopes 6, 4, and 3, of the corresponding curves. 
Also shown is the distribution of size in the case of the destruction of random 
clusters ($\diamond$), with full line indicating the slope 2.5. (b) The distribution $P(\Delta t)$ of 
waiting time $\Delta t$ between fragmentation events, shown with full symbols, and the 
distribution of waiting 
times of individual units involved in the events, empty symbols, for same 
values of $p$ and the fragmentation of the maximal cluster. Large data are logarithmically binned.}
\label{fig-aRGall}
\end{figure*} 
\end{widetext}

Starting with $N=1000$ units, we apply the rules of random aggregation and 
fragmentation events, as described above. 
If we always select the largest cluster for fragmentation, instead of a randomly 
selected cluster, it affects the system dynamics in a different ways, depending on 
potential size that the cluster can grow, which, on the other hand is determined 
by the fragmentation probability $1-p$. For a sizable probability of fragmentation, 
the system has no large clusters, which causes the probability distribution of 
cluster sizes to decay rapidly. The distribution of all sizes of clusters and the size 
of 
the fragmented cluster is shown in Fig.\ \ref{fig-aRGall}d for several values of 
the probability $1-p$ = 10\%, 5\%, and 2\%. It exhibits a power-law tail for sizes above a threshold value $s> s_0$ with the scaling exponent $\tau $ defined by the expression 
\begin{equation}
P(s) \sim s^{-\tau },  s > s_0 .
\label{eq-psize}
\end{equation}

In each case, the size of the largest 
(fragmented) cluster starts playing a dominant role over a threshold $s_0$ 
(maximum of the fragmented size distribution), which moves upward with reduced 
fragmentation probability. The slope $\tau $ of the cluster-size distribution over the 
threshold is fully determined by the distribution of the fragmented cluster size 
and it is continuously varying with the probability $p$.   With the reduced 
fragmentation 
probability (increasing aggregation probability $p$) the slope of the size 
distribution reduces and converges towards the universal curve obtained in the random cluster fragmentation (top line in Fig.\ \ref{fig-aRGall}d). Intuitively, this is understandable since, for vanishingly small but finite probability of crush, a very large clusters may occur in the system increasing the probability to be selected by random picking unit. 
However, the nature of correlations that the system develops over time reveals that these are different processes. The time fluctuations in the number of clusters and their 
power-spectra are shown in Fig.\ \ref{fig-aRGall}a,c for fixed aggregation probability. Apart from the  visual difference, in particular in the sizes of destruction events 
and the 
average number of clusters present (697 in random, 628 in maximal cluster 
destruction events), the power spectrum of these time series shows 
long-range correlations with different exponents. Specifically,  the power spectrum $S(\nu )$ decays at high frequencies $\nu$ as 
\begin{equation}
S(\nu) \sim {\nu}^{-\phi},
\label{eq-spectrum}
\end{equation}
with the scaling exponent $\phi$ depending on the fragmentation mechanism: We find the  exponent close to $\phi \sim 1.6$ for the maximal cluster fragmentation, compared 
to the short-range correlations in the case of the random cluster fragmentations 
$\phi \sim 1.9$ and same aggregation probability $p=0.9$. 

Further insight into the mechanisms of the cluster dynamics can be gained 
from the temporal statistics, in particular, the distribution of the time 
intervals between events. This distribution can be looked at the level of the 
whole system, as well as the level of each unit. The results of the simulations 
are shown in Fig.\ \ref{fig-aRGall}b. The distribution between the 
fragmentation events is exponential, reflecting the fact that in the model the 
fragmentation occurs randomly with the external parameter $1-p$. However, from the 
point of view of the individual units, the distribution of the time intervals 
between two successive fragmentation events in which a unit has been involved 
is rather non-trivial. 
The distribution seems to be primarily determined by the topology (a random annealed 
graph in this case) where the linking between the units occur, rather than with the probability of 
fragmentation (see also discussion in the next section).

\section{Cluster dynamics on networks with fixed and emergent topology \label{sec-networks}}
	
\begin{figure}[htb]
\begin{center}
\begin{tabular}{cc} 
\resizebox{16pc}{!}{\includegraphics{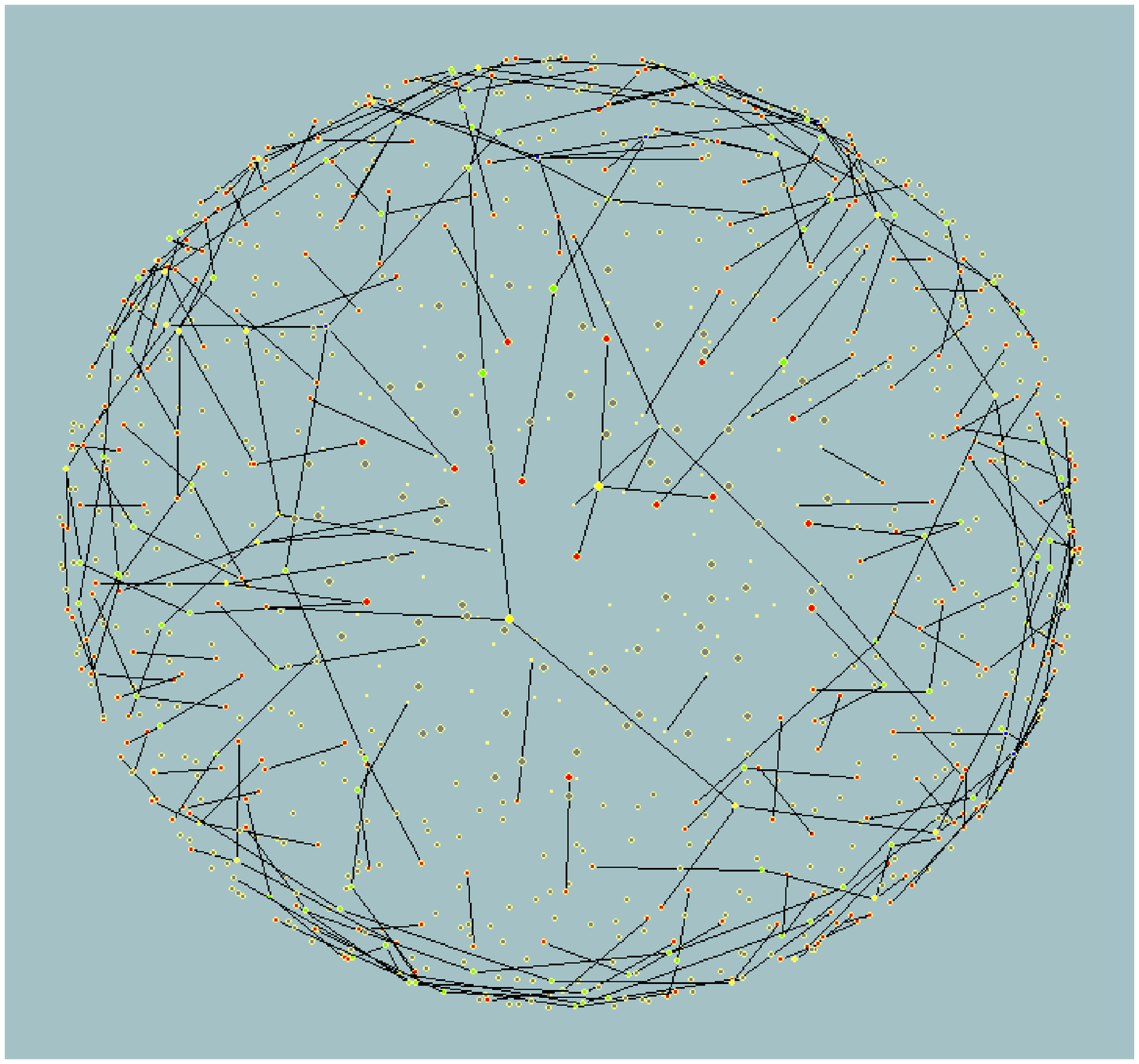}}\\
 \resizebox{16pc}{!}{\includegraphics{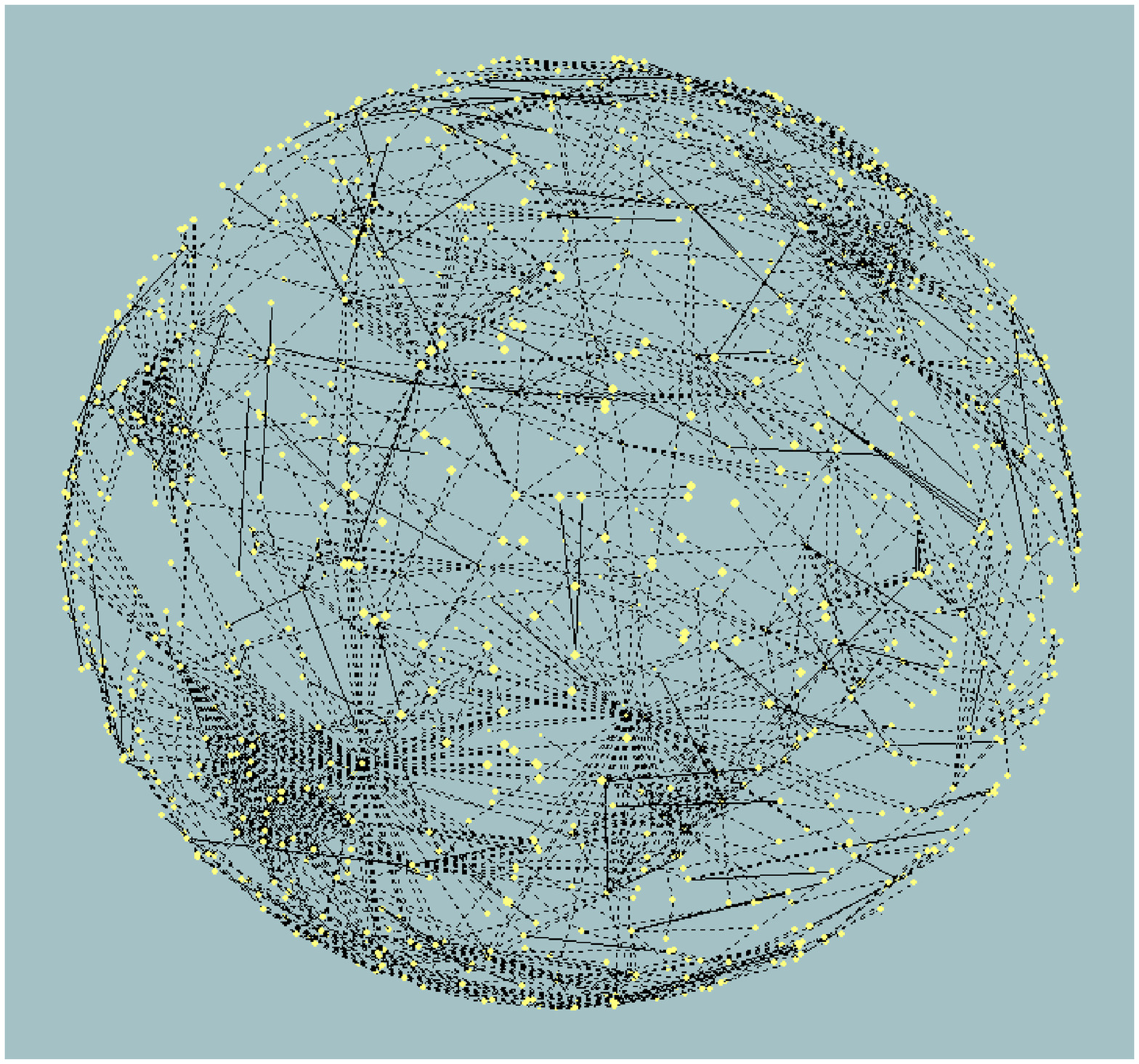}} \\
\end{tabular}
\end{center}
\caption{Emergent networks in two types of cluster dynamics: (top) annealed random graph
 topology, and (bottom) network built via action events on the tree-of-trees structure.}
\label{fig-graphs}
\end{figure}

	In this section we test the effect of placing the aggregation process on a network. 
      In particular,  the units are associated with network nodes, while the network 
	structure permits the aggregation only along the links between the 
	units. Thus, in contrast to the annealed random graph topology discussed 
	above, on the network a cluster can grow by aggregating with locally 
	available units. Apart from the network structure constraints in the 
	aggregation process, the action event occurs with the probability 
	$1-p$ always involving the largest cluster as above. After each action 
	event, the units who were involved in the action can build new 
	connections, specifically,  they form a clique of the size of the
	cluster, which then plays a role in future aggregation. Thus, the 
	dynamical reconstruction of the network structure is tightly linked to the 
	action events. We consider the temporal and statistical features of the 
	process and the structure of the emergent network. Here we present the 
	simulation results starting from a tree network which contains several 
	subtrees with local hubs (the algorithms to grow such networks are 
	described in \cite{MMBT09}).  An example of the network structure emerging 
	through the cluster  is shown in Fig.\ \ref{fig-graphs} along with the 
	network representing the situation of the annealed random graph topology, 
	studied above in sec.\ \ref{sec-model}.	
As shown below, these rules with the aggregation on network and action 
	events with the reconstructed links appear to be a suitable basis for describing the 
	evolution-dependent power-law exponent (which was observed  for example in the 
	analysis of the empirical data of war events in Columbia \cite{johnson}). 
	For comparison, we also consider the case when the network structure remains 
	fixed throughout the entire process. In tis case after the action involving the largest cluster, the links between involved units are always reset to the original network structure.
\begin{figure}[htb]
\begin{tabular}{cc} 
\resizebox{18pc}{!}{\includegraphics{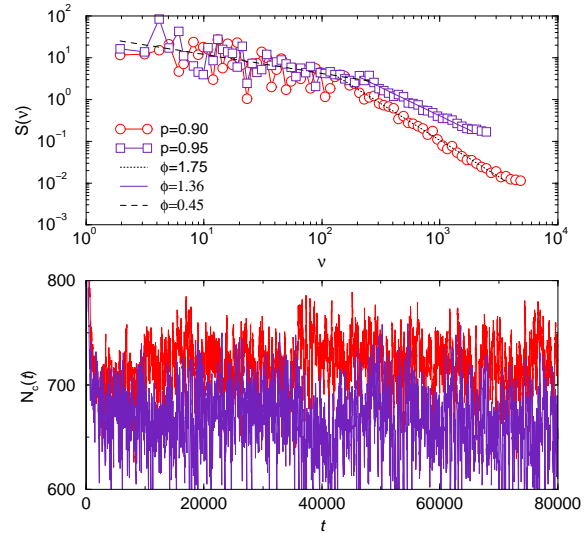}} \\
\end{tabular}
\caption{Time series (bottom) and power spectrum (top) of the number of clusters in the case of maximal-cluster action mechanism on the evolving network for two values of aggregation probability probability $p$.  }
\label{fig-nstETOT}
\end{figure} 

The network reconstruction with building new connections  between units involved in the action is 
affecting the dynamics through increased temporal correlations. In Fig.\ \ref{fig-nstETOT} the example of two time series of number of clusters are shown together with their power spectra, 
corresponding to two values of the parameter $p$. We find the power-spectrum correlations at hight frequencies with the exponent 
$\phi =1.75 \pm 0.026$ for $p=0.9$, while  $\phi =1.36\pm 0.035$ for $p=0.95$.  
Further increase of the correlations towards the $1/f$ noise behavior is expected in the 
asymptotic limit of small action probability. In both cases only weak correlations occur at small frequencies (long times).

\begin{figure}[hbt]
\begin{tabular}{cc} 
\resizebox{18pc}{!}{\includegraphics{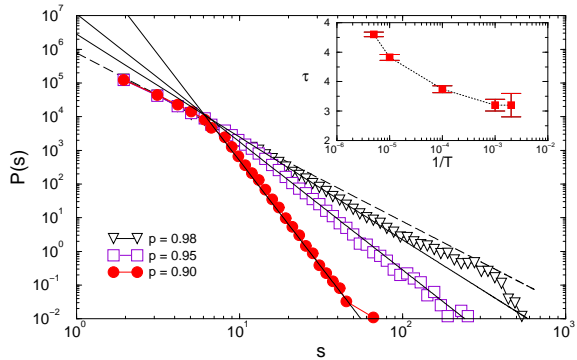}} \\
\end{tabular}
\caption{Size distribution of all clusters on the evolving network for $p=$0.9,0.95, 0.98, and evolution time $T=10^5$ steps. Full lines indicate slopes $\tau =$6.3,3.8, 3.1, while dashed line has the slope $\tau = 2.5$. Inset: Scaling exponent $\tau$ plotted against inverse evolution time $T$ for fixed $p=0.95$.
}
\label{fig-psnet}
\end{figure}

The distribution of cluster sizes, shown in Fig.\ \ref{fig-psnet}, also varies with 
$p$ with the tail slope dominated by the distribution of the maximal cluster 
involved in the action. The exponents in the case of the emergent network built 
on the tree-of-trees structure are found to be smaller compared to the maximal 
cluster sizes on the annealed random graph. Specifically, the exponents are 6.3, 
3.8, and 3.1 for $p=$0.9, 0.95 and 0.98, respectively. The results are for the 
evolution time $T=10^5$ steps. Simulations for different evolution times lead 
to different scaling exponents, as shown in the inset to Fig.\ \ref{fig-psnet}. 
Within our model the temporal dependence of the exponent $\tau$ is related to 
the evolving network structure. Specifically, the network is reconstructed by 
gaining new links after each action event, as explained above, thus the average connectivity 
of units increases over time. In the average, the number of the 
action events for the evolution time $T$  is given by $(1-p)T$.    The increased connectivity
 enables larger clusters to form, however, this occurs at the 
expence of smaller clusters. Consequently, the distribution is steeper for longer evolution time. 
 The decrease of the exponent with decreasing 
evolution time can be also understood by considering the aggregation probability as 
follows: In order to have the same network structure statistically when the evolution 
time $T$ is doubled, $T\to 2T$, one needs to reduce the number of 
the reconstruction events per time step, i.e., $1-p \to (1-p)/2$. This means 
that the aggregation probability is effectively larger and thus the 
exponent closer to 5/2.

\begin{figure}[hbt]
\begin{tabular}{cc} 
\resizebox{16pc}{!}{\includegraphics{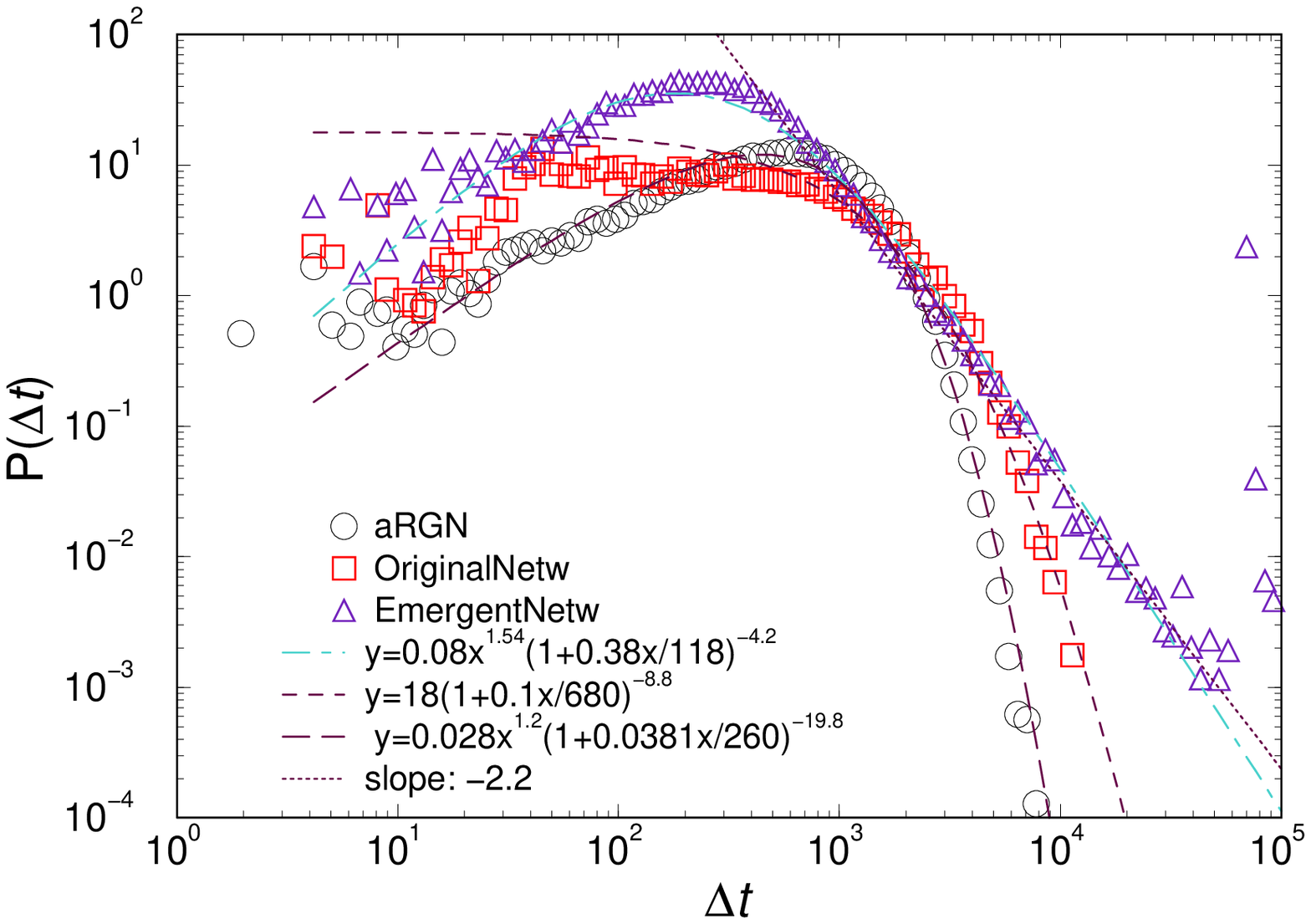}}\\
\resizebox{16pc}{!}{\includegraphics{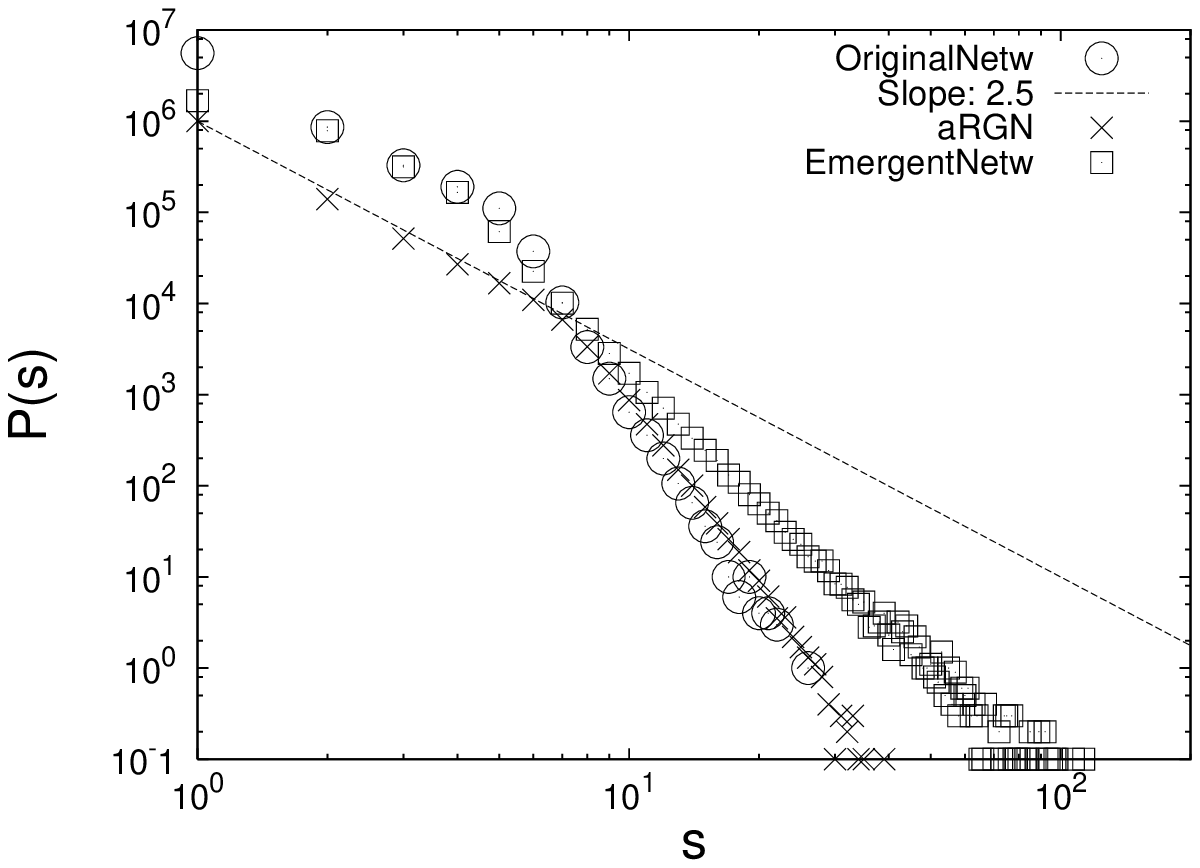}} \\
\end{tabular}
\caption{(top)Distribution of time intervals between consecutive action events of 
a particular unit, averaged over all 1000 units, for three 
network topologies: annealed random graph,  fixed network of tree-with-subtrees, 
and evolving network built on the tree-with-subtrees.  (bottom) Distribution of sizes of all clusters for 
the same networks. Fixed aggregation probability $p=0.9$ and simulation time $T=10^4$ steps.}
\label{fig-3nets}
\end{figure} 

Further comparative analysis of the network effects on the cluster dynamics is given 
in Fig.\ \ref{fig-3nets}. Here we also consider the case where the underlying tree 
network remains fixed in the actions, i.e, following the action events only the original links are preserved. It is remarkable that, when 
the largest cluster starts dominating the dynamics (above the threshold value 
$s_0 \approx 6$ in this case), the tail of the distribution on the tree network 
coincides with the tail obtained in the annealed random graph topology. 
This supports our conclusion that the local structure of the network plays 
a pivotal role in this cluster dynamics. Specifically, the tree graph structure 
and the annealed random graph, which is locally tree-like, have the 
same effects statistically on the cluster aggregation and consequently on 
the size of the largest cluster.

Apart from the cluster size distribution, 
we find that the network  structure also plays a role in how often the units 
participate in the action. The distribution of waiting times between successive 
actions of a unit, averaged over all units on the network, is shown in 
Fig.\ \ref{fig-3nets}(top) for the three network topologies. Here we see 
differences appear between annealed random graph and fixed tree network.
Apart from small times, the data can be fitted  by the $q-$exponential function (see  \cite{tsallis08} for the origin of this distribution in long-range interaction systems): 
\begin{equation}
P(\Delta t) = B(\Delta t)^\alpha[1+(1-q)\Delta t/T_0]^{1/(1-q)} \ ,
\label{eq-qexp}
\end{equation} 
with a prefactor and different parameters. Specifically, $q=$1.04 (exponential limit!) for the random graph topology, and 
$q=$1.12 for the fixed tree network. Whereas, a prominent power-law tail  with the with the slope $\tau =$2.66, compatible with $q=$1.38, occurs in the case of the evolution of network. This implies that the units which are deployed less often than the threshold time ($T_0 \sim 10^3$ steps), appear to be self-organized  in a hierarchical manner.

\section{Conclusions\label{sec-conclusions}}

We have studied the cluster aggregation--fragmentation processes on networks and 
assuming that the {\it largest available cluster is always involved in the 
fragmentation or action 
events}, which also may affect the network structure. This is motivated by the dynamics of conflicts, where planned war actions deploy the largest forces (and may result in correspondingly large damage). 
The underlying network with its fixed or dynamically evolving structure
 represents constraints for the aggregation processes, thus affecting the largest 
cluster sizes, which is relevant for war conflicts. We also studied the situation 
where such constraints do not exist (e.g., in the herding events in stock market 
dynamics, where the information is globally available), resulting in an annealed 
random graph structure. 
 The process is controlled by the aggregation probability $p$, which may 
result from another coupled stochastic process, but here it is taken as an 
external parameter.

We find a number of new features, in both the geometric and temporal patterns 
of the process, which can be related to the network topology. In 
particular, the system is characterized by:
 \begin{itemize}
 \item non-universal power-laws in the distributions of the sizes of events; the 
scaling exponents appear to vary with the fragmentation probability and the 
evolution time (in the case of the evolving network);
 \item long-range temporal correlations in time series representing fluctuations 
of the number of clusters; the correlations are intensified on  networks whose 
structure evolves through the action events;  
 \item evolution of the internal organization, i.e., connectivity and hierarchy 
between units, as a result of the action events.
 \end{itemize}

The distribution of cluster sizes in the mechanism when the largest
 cluster is always selected for the action, is conditioned by the local 
connectivity of the network. Aggregation process on trees yields large number 
of small clusters due to sparse connections on the tree. Lack of large clusters 
results in the large scaling exponent of the size distribution. Within network 
terminology, the cluster aggregation--fragmentation processes in ``free space'' 
leads to a sparse random graph structure with annealed links. Its local structure 
is also tree-like, leading to statistically similar cluster size distribution as 
on the fixed tree graphs. On the other hand, on the evolving network which gains 
new links (cliques) after the action events, the node connectivity  is 
progressively  increasing. This allows larger clusters to form, and consequently, 
smaller scaling exponent of the size distribution, compared to the sparse networks 
for the same fragmentation probability.
In both cases the exponent decreases when the fragmentation probability 
$1-p $ is reduced. In the limit $p\to 1$ with a vanishingly small but finite 
fragmentation probability, the distribution is expected to approach the 
classical slope 5/2 of random aggregation--fragmentation processes, however, 
the cutoff might be different. This limit needs more theoretical consideration, 
which is left out of the present work. 
The relation between the network connectivity and the role of the largest 
cluster in the dynamics is also the basis for the observed variation of the 
scaling exponent of all clusters $\tau$ with the evolution time of the network in our model. For a 
fixed fragmentation probability (number of reconstruction events per time step), the larger 
evolution time leads to larger size of the fragmented cluster. Hence the  distribution of all 
clusters is steeper, compared to  small evolution time, where the difference between cluster 
sizes is reduced and thus closer to the processes with random cluster events.

 Our analysis of temporal features of the system reveals the nature of the 
process underlying the observed cluster size distribution. Namely, systematic 
network reconstruction following the actions over time leads to a complex 
organization between units which causes them to play different roles 
(frequencies) in the actions. Despite the temporal correlations, which are 
caused by the involvement of the largest cluster in the events,  such internal 
structure does not occur in the original aggregation--fragmentation model in 
free space and locally sparse networks. Thus the network reconstruction via the 
cluster dynamics proposed in our work provides the mechanism for building an 
effective structure of units, which has large capacity (in terms of the size of 
events) and a self-organized internal structure.

\acknowledgments
We are grateful to  Collaborative Linkage Grant CBP.NR.NRCLG.982968 for partial financial 
support of this work. We would like to thank V. Priezzhev for helpful discussions. 
B.T.  also acknowledges support from the program No. P1-0044 (Slovenia) and COST-MP0801 action ``Physics of competition and conflicts''.


\begin{thebibliography}{99}

\bibitem{wattis} J. A. D. Wattis, Physica D {\bf 222}, 1 (2006).
\bibitem{smol} M. von Smoluchowski, Z. Physik  {\bf 7}, 557 (1916).
\bibitem{becker} R. Becker and W. Doring, Ann. Phys. {\bf 24}, 719 (1935).
\bibitem{bouchaud}R. Cont and J.P. Bouchaud, Macroeconomic Dynamics {\bf 36}, 394 (2000).
\bibitem{EZ} V.M. Eguiluz and M. G. Zimmermann, Phys. Rev. Lett. {\bf 85}, 5659 (2000). 

\bibitem{geoff1}	R. Dhulst and G. J. Rodgers, Int. J. Pure and Applied Finance {\bf 3}, 609  (2000). 

\bibitem{johnson}	N. Johnson, M. Spagat, J. Restrepo, J. Bohorquez, E. Restrepo and R. Zarama, 	arXiv:physics/0605035.
\bibitem{richardson48} L. F. Richardson, Amer. Stat. Assoc. {\bf 43}, 523 (1948).
\bibitem{richardson60}L. F. Richardson, Statistics of Deadly Quarrels, eds. Q. Wright and C. C. Lienau 	(Boxwood Press, Pittsburgh, 1960). 
\bibitem{newman}M. E. J. Newman, Contemp. Phys. {\bf 46}, 323 (2005).
\bibitem{clauset} A. Clauset and M. Young, physics/0502014.

\bibitem{rose} B. Rosenberg and and J. A. Ohlson, J. Fin. Quant. Anal. {\bf 11}, 393 (1976).

\bibitem{MMBT09} M. Mitrovi\;c and B. Tadic, Phys. Rev. E  {\bf 80}, 026123 (2009).
\bibitem{tsallis08}A. Pluchino, A. Rapisarda, and C. Tsallis, arxiv: 0809.4850.


\end{thebibliography}
\end{document}